\documentclass[trackchanges,twocolumn]{aastex7}
\usepackage{amsbsy}
\usepackage{amsmath}
\usepackage{comment}
\usepackage{xspace}
\usepackage{multirow}
\usepackage{booktabs}

\newcommand{\msun}{\mbox{M$_\odot$}}

\newcommand{\kpc}{\mbox{${\rm kpc}$}}

\newcommand{\jwst}{\textit{JWST}\xspace}
\newcommand{\hst}{\textit{HST}\xspace}
\defcitealias{harris_reinacampos2023}{Paper I}
\defcitealias{harris_reinacampos2024}{Paper II}

\begin{document}

\title{On the correlation between globular clusters and the distribution of dark matter in galaxy clusters: the case of Abell 2744}

\author[orcid=0000-0002-8556-4280]{Marta Reina-Campos}
\affiliation{Canadian Institute for Theoretical Astrophysics (CITA), University of Toronto, 60 St George St, Toronto, M5S 3H8, Canada}
\affiliation{Department of Physics \& Astronomy, McMaster University, 1280 Main Street West, Hamilton, L8S 4M1, Canada}
\affiliation{Instituto Galego de F\'isica de Altas Enerx\'ias, Universidade de Santiago de Compostela, 15782 Santiago de Compostela, Galicia, Spain}
\email[show]{marta.reina@usc.es}

\author[orcid=0000-0003-2573-9832]{Joshua S.~Speagle}
\affiliation{Department of Statistical Sciences, University of Toronto, 9th Floor, Ontario Power Building, 700 University Avenue, Toronto, ON M5G 1Z5, Canada}
\affiliation{David A. Dunlap Department of Astronomy \& Astrophysics, University of Toronto, 50 St George Street, Toronto, ON M5S 3H4, Canada}
\affiliation{Dunlap Institute for Astronomy \& Astrophysics, University of Toronto, 50 St George Street, Toronto, ON M5S 3H4, Canada}
\affiliation{Data Sciences Institute, University of Toronto, 17th Floor, Ontario Power Building, 700 University Avenue, Toronto, ON M5G 1Z5, Canada}
\email{j.speagle@utoronto.ca}

\author[orcid=0000-0001-8762-5772]{William E.~Harris}
\affiliation{Department of Physics \& Astronomy, McMaster University, 1280 Main Street West, Hamilton, L8S 4M1, Canada}
\email[show]{harris@physics.mcmaster.ca}

\begin{abstract}
Globular clusters (GCs) lie scattered around the inner $40\%$ of the virial radius of galaxy clusters, potentially  being excellent tracers of the underlying mass distribution. In this paper, we present a statistical method based on assuming that the location of GCs around a galaxy cluster follows an inhomogenous spatial Poisson point process, and we use this method to assess to which galactic component GCs are better tracers of. We apply the method to the galaxy cluster Abell 2744, and we find that the spatial distribution of bright GCs roughly traces the three main interacting clumps in the galaxy cluster, alongside other galaxies with sizeable GC populations. The GC populations are more closely correlated to the predicted mass maps than any other galactic component (Spearman rank coefficients $>0.7$). A perk of this statistical method is that it allows us to distinguish to which map the agreement is closest to. In particular, we find that the Bright Blue GCs are compatible with the mass map solely derived from weak lensing, suggesting that they can provide complementary and independent information on the mass distribution in galaxy clusters with a similar level of detail to that of weak lensing. This statistical method is available in a public repository, and combined with  catalogs of GCs in galaxy clusters at different cosmic epochs, it provides an independent method for investigating the mass distribution in these galactic environments. 
\end{abstract}



\section{Introduction} \label{sec:intro}

The nature of dark matter (DM) is one of the main unsolved problems in physics. As the dominant component of mass in the Universe \citep{planck2018}, it is a crucial ingredient in galaxy formation. Because DM cannot be directly probed using stellar light, the most direct way to determine its presence and nature must be from their gravitational influence on other objects. 

One of the best environments to probe the microphysics of DM is around massive galaxy clusters \citep[e.g.][]{banerjee20,asencio21}, which have long been studied through gravitational lensing \citep[see reviews by][]{treu10, kneib11, hoekstra13}. Their large masses, combined with their distance from us, lead to background galaxies being stretched and magnified (i.e.~strong lensing) or having small deformations in their shapes (i.e.~weak lensing). By modeling the size and location of the structure producing these effects, the DM distribution within the galaxy cluster can be reconstructed. Additionally, recent studies complement lensing effects with spectroscopically-selected galaxy members to construct robust high-resolution mass maps \citep[see][]{grillo15}. 

Building a detailed mass distribution of the galaxy cluster requires a large amount of observational data, from deep imaging to identify multiply-lensed images to spectroscopic campaigns to confirm the redshift of these images. In this respect, the launch of the \textit{James Webb Space Telescope} (\jwst) has been revolutionary. By performing deep, wide and multiband observations of lensing galaxy clusters, high-resolution maps of the mass distribution in tens of galaxy clusters are being constructed \citep[e.g.][]{furtak23,diego23,diego23b,diego24b,gledhill24,cerny26, diego25,price25,rihtarsic25,allingham26}. However, the challenge of identifying the nature of DM is multi-faceted, and independent observational proxies to trace the DM distribution in galaxies are needed to provide complementary views.

An alternative observational tracer of the potential well of the galaxy cluster is through X-rays, which are emitted through thermal bremmstrahlung by the highly ionised Intracluster Medium (ICM) gas bound to the cluster, and the emission is directly linked to the total gravitating mass contained in the cluster, so it can be efficiently used as a tracer of their mass. However, the collisional (dissipative) nature of gas implies that it will experience ram pressure when galaxies collide, producing an offset between the X-ray emission and the location of the DM haloes. A classical example of this effect is the well-known Bullet Cluster \citep[e.g.][]{clowe04, markevitch04,cha25,cho25}. The use of X-ray emission as a tracer of DM mass is thus restricted to relaxed galaxy clusters \citep[e.g.][]{montes19}, which are the minority of cases. For this reason, exploring additional alternative luminous tracers for the detailed mass distribution is imperative to complement the constraints from gravitational lensing.


In the current $\Lambda$CDM cosmological paradigm, galaxy clusters hierarchically assemble by accreting galaxies or small galaxy groups. A revealing signature of this process is the intracluster light \citep[ICL;][]{mihos16,montes22b}, i.e.~free-floating stars not bound to any galaxy in the cluster. Similarly, this assembly process deposits globular clusters (GCs) in the intracluster medium up to $40\%$ of the virial radius \citep[e.g.][]{harris+2020, kluge25}. Because the intracluster light and GCs are relics of the assembly process of the galaxy cluster, their physical scales are comparable to those of the DM distribution \citep{dubinski98}. A key aspect of these visible tracers is that their collisionless nature leads them to follow the distribution of mass in galaxy collisions, in contrast to the dissipative gas, which drags behind. Thus, the ICL and GCs are good candidates for being visible tracers of the spatial distribution of mass around the galaxy cluster, specially in non-virialised galaxy clusters~\citep{montes19}.

An additional key aspect of using GCs is that their high surface brightness and compact size, coupled to the effects of stellar evolution and photometric corrections, make them as identifiable at moderate redshifts as they are in the Local Universe with appropriately designed observations  \citep[see discussion in][]{harris25a}. The observations of lensing galaxy clusters by \jwst at intermediate redshifts ($0.2 < z < 1$) are allowing us to see GCs and ultra compact dwarf (UCD) populations at different evolutionary stages \citep[some early studies include][]{faisst+2022,lee22,diego23,harris_reinacampos2023,harris_reinacampos2024,martis24,keatley25,berkheimer26,diego26a,hinrichs26}\footnote{Most of the photometric catalogues of GCs produced in these studies are collected in the RESCUE website: \href{https://mreinacampos.github.io/starclusters-in-jwst/}{https://mreinacampos.github.io/starclusters-in-jwst/}}. These catalogues provide an excellent raw material to statistically examine the question: \emph{`Does the spatial distribution of GCs trace that of DM?'} and \emph{`Is there another galactic component that traces DM at a similar precision?'}.

A growing body of evidence explores the correlation between the spatial distribution of GCs and that of DM in galaxy clusters, both in \textit{HST} and \textit{JWST} observations \citep[e.g.][]{alamo-martinez13,lee_jang2016,lee22,diego23,diego24b,martis24,diego26a}, and in numerical simulations \citep{reina-campos+2022,reina-campos+2023}. In most of these relaxed systems, the correlation is examined by comparing the 1D projected radial distribution of GCs and the inferred mass in the galaxy cluster, finding good agreement between both, though the GC profile tends to be steeper. However, this technique does not work in non-spherical or non-relaxed galaxy clusters with multiple interacting brightest cluster galaxies (BCGs). Thus, a new method to infer the correlation between the spatial distribution of GCs and of matter in galaxy clusters is needed that will able to deal with more complex 2D distributions that may not have any particular symmetry.

In this paper, we introduce a statistical method to assess whether the spatial distribution of GCs in a galaxy cluster is a good tracer of the underlying mass distribution, or of any other galactic component. For this, we use the population of GCs in the galaxy cluster Abell 2744 \citep[][hereafter Paper I and Paper II, respectively]{harris_reinacampos2023,harris_reinacampos2024}, and we compare their distribution against the predicted mass distribution from strong gravitational lensing, from the observed stellar light, and from the X-ray emission. This galaxy cluster is located at a redshift of $z=0.308$, and it is a massive, X-ray luminous, merging galaxy cluster \citep{allen98, ebeling10}. Nicknamed the `Pandora cluster', it has been the target of extensive multiwavelength observations, including ultradeep \jwst observations by the UNCOVER Treasury survey \citep{bezanson24} and the MegaScience Cycle 2 survey \citep{suess24}. As one of the most powerful known gravitational lensing clusters, it has a large area of high magnification because it consists of several merging subclusters \citep[e.g.][]{merten+2011,richard14,wang15,jauzac15,diego16,kawamata16}. Thus, it is an excellent and challenging test bed for our statistical method for comparing GCs and the underlying mass in non-relaxed galaxy clusters.

We describe the statistical method to compare the distributions in Sect.~\ref{sec:methods}, the data in Sect.~\ref{sec:data} and a qualitative description of the maps in Sect.~\ref{sec:results}. We then apply our method in Sect.~\ref{sec:probabilities}, and conclude with a discussion on future prospects in Sect.~\ref{sec:conclusions}.

\section{Methods} \label{sec:methods}

As stated avove, the goal of this paper is to address two questions: \emph{`Does the spatial distribution of GCs trace that of DM?'} and \emph{`Is there another galactic component that traces DM at a similar precision?'} in the context of the galaxy cluster Abell 2744. For that we require an statistical framework to compare the spatial distribution of bright star clusters against different  distributions, which we motivate here\footnote{All the code related to this project is publicly available here: \url{https://github.com/mreinacampos/starclusters-in-jwst/tree/main/code_poisson_comparison/abell2744}}.

\subsection{General method: inhomogeneous Poisson point process}

To carry out this comparison, we assume that the location of bright star clusters around a galaxy cluster can be modeled with an inhomogeneous spatial Poisson point process\footnote{See also \citet{li2025} for another use case of modelling GCs using point processes.}. This process has an intrinsic occurrence rate that varies across the field of view, $\lambda(x,y)$, which we parametrize in terms of the pixels along the horizontal and vertical axes of projected plane, $(x,y)$, respectively. We assume that the occurrence rate $\lambda$ is represented by different images of galactic components in the galaxy cluster, and thus we use the same notation for describing the intrinsic rate and the observational maps.

Because we are modeling the probability of \emph{observing} star clusters scattered around a galaxy cluster, we need to define an effective occurrence rate,
\begin{equation}
    \lambda_{\rm eff}(x, y, \pmb{\theta}) = \lambda(x, y) S(\pmb{\theta}),
\end{equation} 
that accounts for the observational probability of observing a given star cluster, $S(\pmb{\theta})$\footnote{Bold font indicates arrays.}. This probability may depend on a number of variables (which might include measurables such as magnitude, color, local sky brightness, crowding level), which we generically describe using the notation $\pmb{\theta}$. 

Given an observed sample of star clusters with data $(x,y,\pmb{\theta})$, we model the log-likelihood that their spatial distribution is given by the effective occurrence rate $\lambda_{\rm eff}$ in the parameter space $B = (x, y, \pmb{\theta})$ of $i = 1,...,k$ bins as,
\begin{align}
    \ln\{\mathcal{P}[N(B_i) = n_i|\lambda_{\rm eff}]\} &= -\sum_{i=1}^{k}\Lambda_{\rm eff}(B_i) - \sum_{i=1}^{k} \ln[n_i!] \nonumber \\ 
    &+ \sum_{i=1}^{k}\ln[\Lambda_{\rm eff}(B_i)]\cdot n_i,
\end{align}
where $n_i$ is the number of observed star clusters in the bin $i$. Under a continuous approximation, i.e.~in the limit when each pixel would contain either 1 or no GC, the log-likehood simplifies to, 
\begin{equation}
    \ln\{\mathcal{P}[(x,y,\pmb{\theta})|\lambda_{\rm eff}]\} \approx  
     \sum_{j=1}^{N_{\rm GCs}}\ln[\Lambda_{\rm eff}(x_j, y_j, \pmb{\theta}_j)],
\end{equation}
where the effective intensity measure is measured at the location in parameter space of each star cluster $j$, $\Lambda_{\rm eff}(x_j, y_j, \pmb{\theta}_j) = \lambda(x_j,y_j)S(\pmb{\theta}_j)$. For comparisons against the same effective rate, the normalization term can be disregarded.

\subsection{Application to Abell 2744}

The rich galaxy cluster Abell 2744 provides an intriguing opportunity to test the methodology developed above.  It does, however, confront us with a fairly challenging test because the cluster is an actively merging systems with three major subcenters and no fewer than five BCGs at the centers of these three subclusters.

To apply the above method to a given population of star clusters, we require a description of their observational probability, and of the quantities affecting such probability. In Abell 2744, the probability of observing star clusters  \citepalias{harris_reinacampos2023, harris_reinacampos2024} can be modelled in terms of a logistic regression function,
\begin{equation}
    S(F150W, \log_{10}\sigma_{\rm sky}) = \left[1+\exp({-g})\right]^{-1},\label{eq:selection-function}
\end{equation}
that depends on the magnitude in the F150W filter, and the logarithmic local sky noise, $\log_{10}\sigma_{\rm sky}$, as,
\begin{equation}
    g = \beta_0 + \beta_1 F150W + \beta_2 \log_{10}\sigma_{\rm sky}.
\end{equation}
The best-fitting parameters are $\beta_0 = 85.84$, $\beta_1 = -2.59\pm0.04$, and $\beta_2 = -5.37\pm0.15$ \citepalias{harris_reinacampos2024}.
This is also called the \emph{recovery probability} in the notation of Paper II and \citet{harris_speagle2024}, or more colloquially the completeness fraction of the photometry.

The luminosity function in the $F150W$ filter is well described by a lognormal distribution,
\begin{equation}
    \phi(m) = \dfrac{N_{\rm tot}}{\sqrt{2\pi}\sigma}\exp\left[\dfrac{-(m-m_0)^2}{2\sigma^2}\right],
\end{equation}
where $\phi(m)$ is the completeness-corrected observed number of objects per unit magnitude. By assuming values for the standard deviation $\sigma$ and running a range of restricted two-parameter solutions on the total population over all luminosities, $N_{\rm tot}$, and the turnover magnitude, $m_0$, \citetalias{harris_reinacampos2024} find that the best solution is $(N_{\rm tot}, m_0, \sigma) = (89,920\pm5400, 31.76\pm0.05~{\rm AB~mag}, 1.4~{\rm AB~mag})$ for the objects in the ``Zone 1'' region with the lowest sky noise and deepest photometric limits.




\subsection{Model testing: posterior predictive comparison}\label{subsec:cross-map}

\begin{figure}
    \centering
    \includegraphics[width=\hsize]{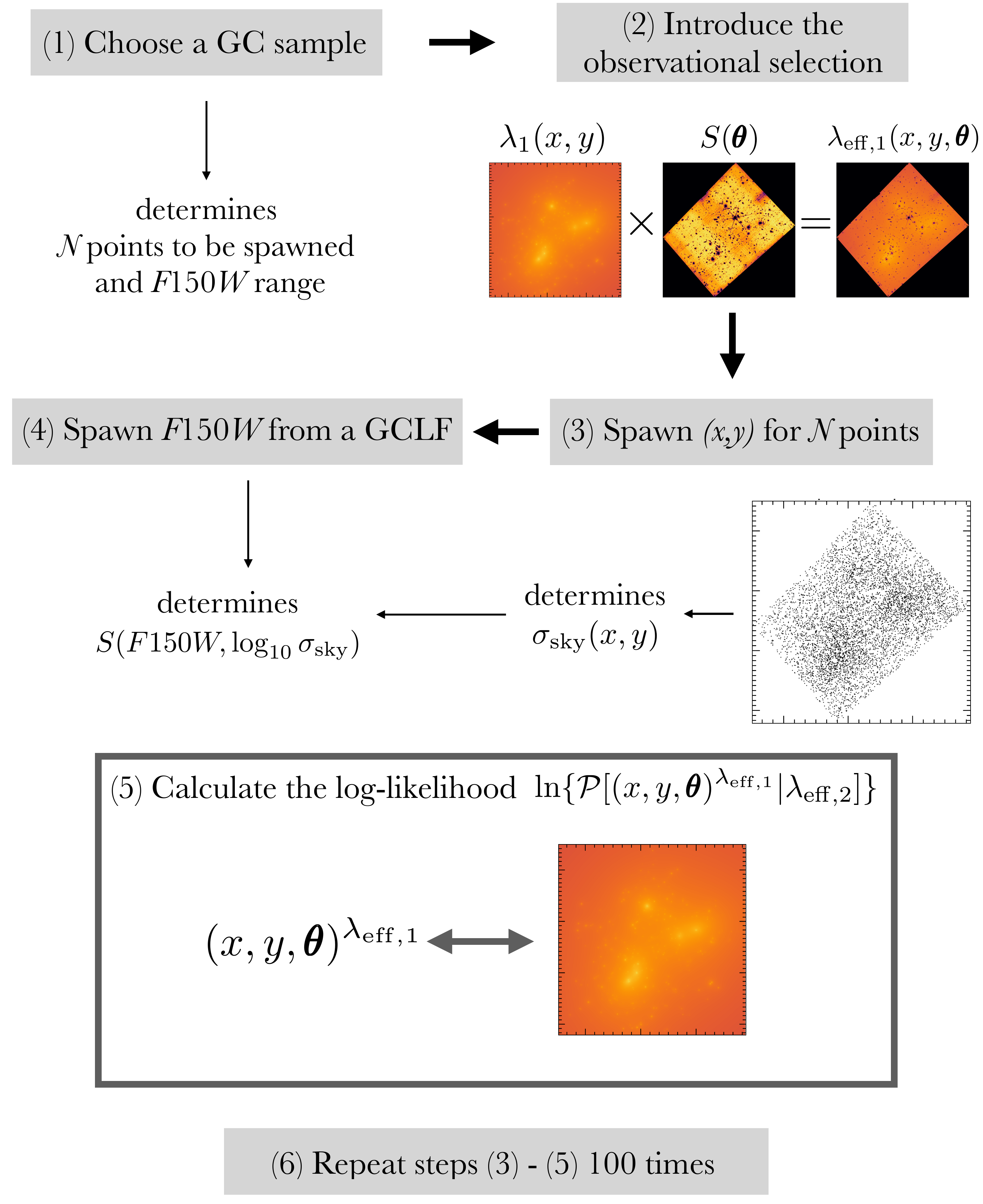}
    \caption{Visual schematic of the cross-map comparison described in Sect.~\ref{subsec:cross-map}: after introducing the observational selection into the map $\lambda_1$ and spawning coordinates for $N$ points, we calculate the log-likehood against a map $\lambda_2$. We repeat the process to obtain a distribution of log-likelihoods.}
    \label{fig:scheme-cross-map-comparison}
\end{figure}

An important element in this comparison is to evaluate the statistical relevance of the log-likelihood of a sample of GCs given an occurrence rate $\lambda$, i.e.~to assess whether a sample of GCs traces a certain galactic component in the galaxy cluster. To do this, we need to test each galactic component by itself by calculating the distribution of log-likelihoods of a component in reproducing itself, which we will use as an anchor in our comparisons. In essence, we are asking: \emph{`Given a predictive occurrence rate $\lambda_1$, how likely is it that points spawned from it can be reproduced with an occurrence rate $\lambda_2$ given an inhomogeneous Poisson point process?'.}

For these tests, we consider two cases. In the first one, both maps are the same, $\lambda_1 = \lambda_2$, and so, we test how well a map can trace itself. We use this `self-map' case as an anchor to assess the statistical relevance of the log-likelihoods. The second case corresponds to two maps that are different. In this dubbed `cross-map' comparison, we are determining the log-likelihood of e.g.~stars or X-ray emission to trace the mass distribution. By comparing these cross-map distributions against the empirical log-likelihoods from the GC samples, we can determine which galactic component is a better tracer of the mass in the galaxy cluster.

To perform this comparison, we adapt slightly the procedure described in Sect.~6.3 in \citet{berek24}. For a pair of maps $\lambda_1$ and $\lambda_2$, the procedure requires spawning the locations $(x,y)$ for a number of points from the map $\lambda_1$, assigning them a realistic magnitude, and comparing the resulting dataset against the map $\lambda_2$ (summarised in Fig.~\ref{fig:scheme-cross-map-comparison}):
\begin{itemize}
    \item[1.] Given a GC sample, we define the $N$ number of points to be spawned as the observed number of GCs, and the magnitude range in $F150W$ in which the magnitudes of the points will be spawned. 
    \item[2.] We introduce the observational bias into the model comparison by modifying the first map with the observational selection function (eq.~\ref{eq:selection-function}). To do that, we measure the local sky noise across the original mosaics from NIRCam, and the resulting image is re-binned and re-centered to cover the same field of view as the map $\lambda_1$. We calculate the selection function for a given magnitude\footnote{The magnitude dependence factors out when re-scaling the map $\lambda_1$ into a probability map from which to spawn $N$ points.}, and apply it to the map $\lambda_1$, thus creating an effective occurrence rate map, $\lambda_{\rm eff, 1}$.
    \item[3.] We randomly spawn the locations $(x,y)$ of $N$ data points from the $\lambda_{\rm eff, 1}$ map. These locations determine their local sky noise, as measured from the NIRCam mosaics. 
    \item [4.] Assuming the observed luminosity function in $F150W$ and the magnitude range given by the GC sample, we spawn $N$ magnitudes and assign them to the data points. Together with the local sky noise, these values determine the observational probability associated with each spawned data point (eq.~\ref{eq:selection-function}), $S(F150W,\log_{10}\sigma_{\rm sky})$.
    \item[5.] We calculate the log-likelihood that these data points have been spawned by the occurrence map $\lambda_{\rm eff, 2}$, $\ln\{\mathcal{P}[(x,y,\pmb{\theta})^{\lambda_{\rm eff, 1}}|\lambda_{\rm eff, 2}]\}$, under the assumption that they can be modeled with an inhomogeneous Poisson point process.
    \item[6.] We repeat this exercise 100 times, and collect the resulting probability densities, which correspond to the distribution of probability densities. We use the expected value, $E[\ln\mathcal{P}]$ and the dispersion around this value to interpret the empirical log-likelihoods of the GC samples.
\end{itemize}
To carry out the comparisons, we introduce the metric,
\begin{equation}
|\mathcal{Z}| = |(\ln \mathcal{P}[\lambda_1|\lambda_2] - E[\ln\mathcal{P}[\lambda_2|\lambda_2]])| / \sigma_{\ln\mathcal{P}[\lambda_2|\lambda_2]},
\end{equation}
which effectively measures how many standard deviations away a map $\lambda_1$ is from a map $\lambda_2$. In the following discussion, we treat scores of $|\mathcal{Z}| < 5$ as compatible. When doing the comparison against two maps (Sect.~\ref{subsec:cross-map}), we only interpret the $|\mathcal{Z}|$--score as an agreement when both directions agree, $\mathcal{Z}(\lambda_1, \lambda_2) \simeq \mathcal{Z}(\lambda_2, \lambda_1) $.

\subsection{Uncertainties on the empirical $|\mathcal{Z}|$-scores}\label{subsec:robustness}

A last check on the robustness of the method\footnote{Also see App.~\ref{app:robustness}} is to estimate the uncertainties around the empirical $|\mathcal{Z}|$-scores, i.e.~the scores for the samples of GCs. To quantify the dispersion around the log-likelihoods of a sample of GCs to be sampled from a map $\lambda$, $\ln\{\mathcal{P}[(x,y,\pmb{\theta})|\lambda_{\rm eff}]\}$, we bootstrap the analysis 200 times. In the following, we quote the $95\%$ confidence interval as uncertainties around the empirical $|\mathcal{Z}|$-scores.

\section{Data} \label{sec:data}

In this Section, we present the different datasets used in this work to describe the bright star clusters scattered around the galaxy cluster Abell 2744, as well as its predicted mass distributions, the stellar light and the X-ray distribution emitted by hot, ionized gas. As an alternative to the galactic components described below, we also consider a uniform background.

\subsection{GC/UCDs in Abell 2744}

We use the sample of bright star clusters in Abell 2744 presented in \citetalias{harris_reinacampos2023} and \citetalias{harris_reinacampos2024}\footnote{Available upon reasonable request via the RESCUE website \href{https://mreinacampos.github.io/starclusters-in-jwst/}{https://mreinacampos.github.io/starclusters-in-jwst/}}. In order to obtain the intrinsic magnitudes and colors of the star clusters, we apply the following cosmological K-corrections \citep{reina-campos_harris2024}: $K_{F115W} = 0.17$, $K_{F150W} = 0.17$, and $K_{F200W} = 0.42$ (see \citetalias{harris_reinacampos2023}). The corrected magnitudes are indicated by $_0$, and we clean the catalog of outliers that have colors outside the range $-1.2<(F115W-F200W)_0<1.2$. 

Based on this catalog, we define four samples of GCs\footnote{Given the very low estimated fraction of UCDs in the sample (\citetalias{harris_reinacampos2023}), we simplify their nomenclature and refer to it as `GCs'.}, all of which only contain star clusters brighter than F150W$<29.5$. This simple first selection is called `Bright GCs'. We also define a `complete' sample by only selecting clusters in Zones 1 and 2 with low sky noise levels ($\log_{10}\sigma_{\rm sky} < 2.1$, see \citetalias{harris_reinacampos2024}). Finally, we divide the sample based on their  $(F115W-F200W)_{0}$ color; the blue, $(F115W-F200W)_{0}<0$, and red, $(F115W-F200W)_{0}>0$, sub-samples are called `Bright Blue GCs' and `Bright Red GCs', respectively. Out of a total of $10,575$ GCs in the catalogue, there are $6624$ bright GCs in our entire sample, $5403$ complete GCs, and $3610$ and $3014$ in the blue and red sub-samples, respectively. 


\begin{figure}
    \centering
    \includegraphics[width=0.9\hsize]{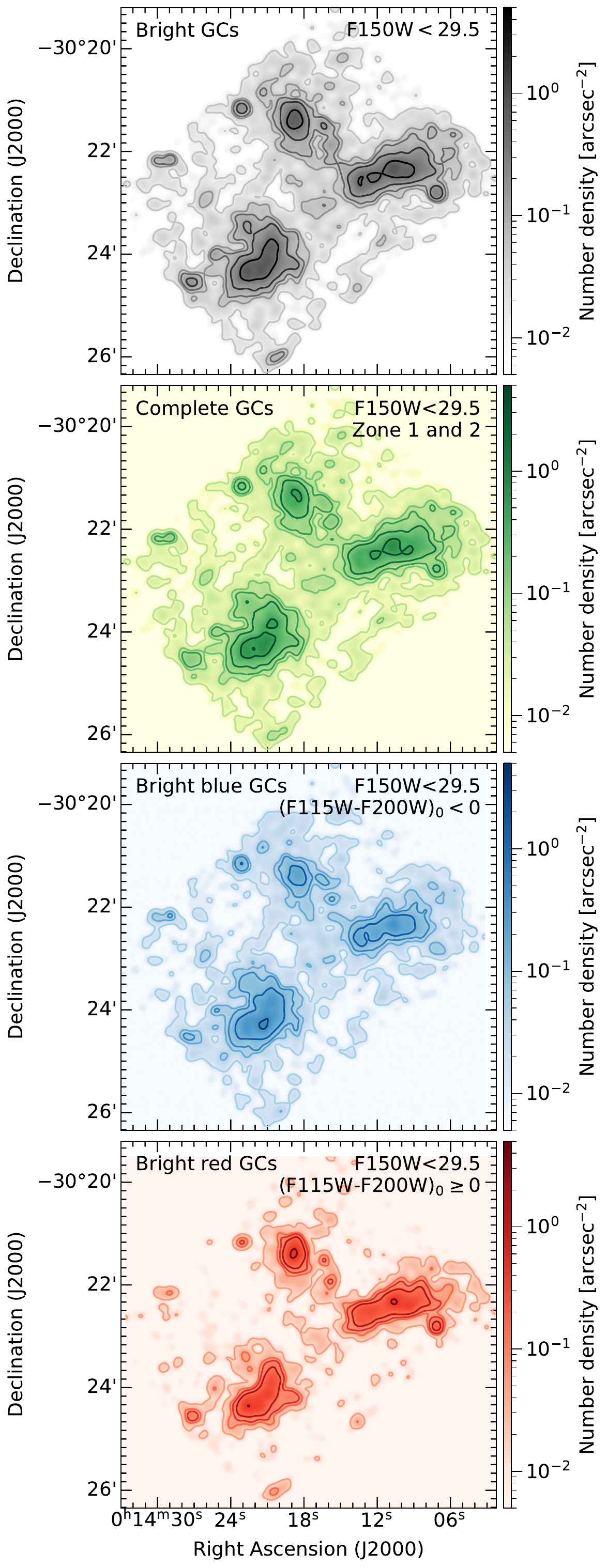}
    \caption{Spatial distributions of GCs brighter than F150W$~<29.5$ in Abell 2744: all GCs (top panel), complete GCs (second panel), as well as blue and red GCs (third and fourth panels, respectively). The number density maps include the observational bias, and are smoothed by a bi-dimensional Gaussian kernel of size $20~\kpc$. Bright clusters are concentrated in three large structures that correspond to the main interacting clumps in the galaxy cluster, with the intracluster medium mostly populated by blue GCs.}
    \label{fig:gcs-contours-all-blue-red}
\end{figure}

\begin{figure}
    \centering
    \includegraphics[width=1\linewidth]{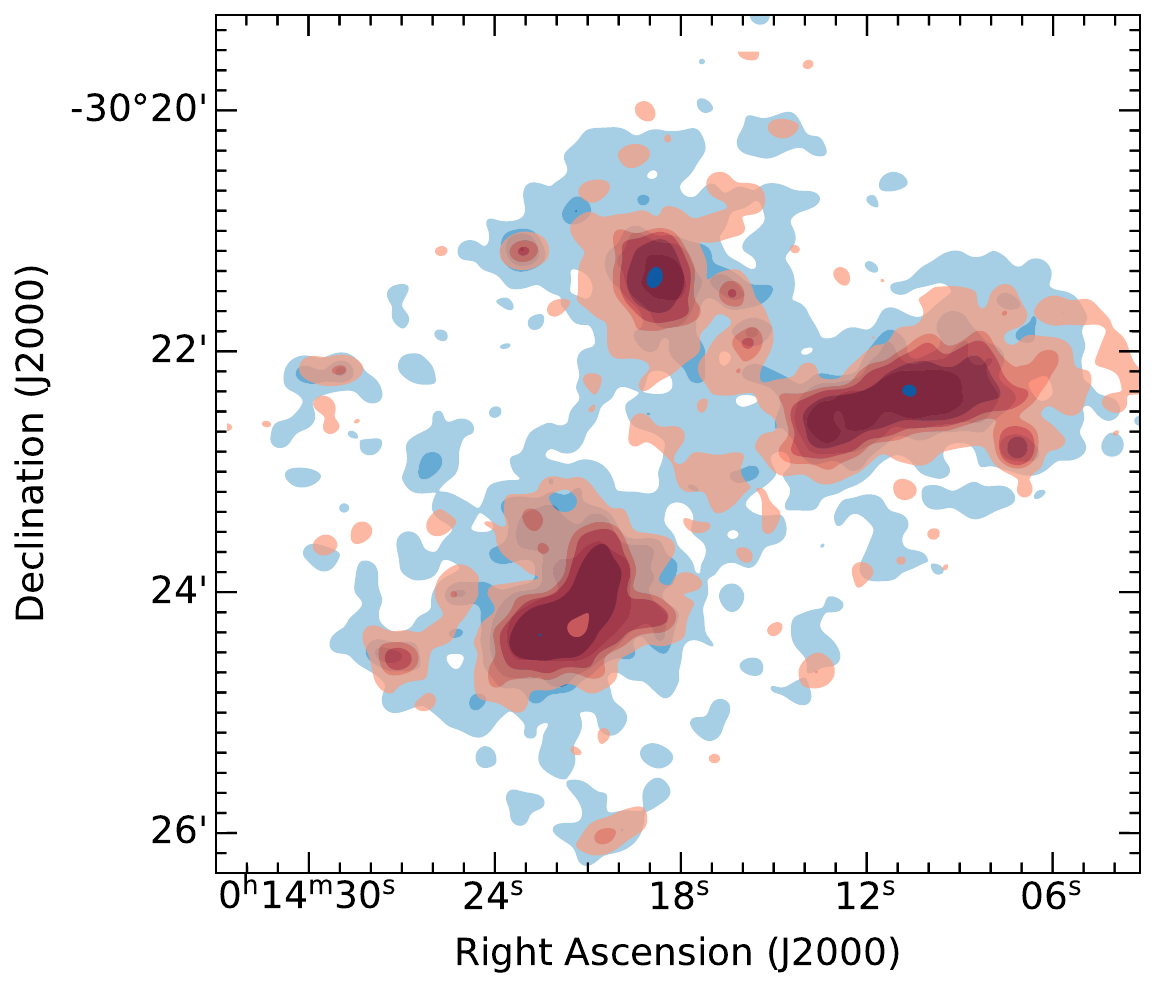}
    \caption{Overlapping spatial distributions of bright blue and red GCs. The filled contours correspond to $\log_{10}\{n/[{\rm arcsec^{-2}]}\} = -1.9, 1.4, -1.2, -0.9, -0.4$. Blue bright star clusters are more extended around galaxies and into the intercluster medium than red ones.}
    \label{fig:gcs-overlapping-blue-red}
\end{figure}

To examine the spatial distributions of GCs, we calculate the number density distribution of each subsample using the \texttt{numpy.histogram2d} routine. The observational bias is included by weighing each cluster $i$ by its probability of recovery, $S_i$. The histograms are then smoothed with a Gaussian kernel of size $20~\kpc$, roughly the effective radii of the galaxies in the galaxy cluster (see the effect of smoothing in the appendix of \citealt{reina-campos+2023}). Using the \texttt{matplotlib.contour} routine, we draw contours enclosing structures with number densities $\log_{10}\{n/[{\rm arcsec^{-2}]}\} = -1.9, 1.4, -1.2, -0.9$, and $-0.4$, which roughly correspond to the $50$th, $80$th, $90$th, $95$th and $99$th percentiles of the distribution of number densities in the smoothed map of blue GCs. These contours allow us to visually identify structures that enclose the same number density of objects. The resulting number density maps are shown in Figs.~\ref{fig:gcs-contours-all-blue-red} and \ref{fig:gcs-overlapping-blue-red}.

We find that the GCs in the four subsamples are mainly clustered in three large structures, at the south-east (`main'), north-west and north of the galaxy cluster. These structures seem to correspond to the three main interacting clumps in Abell 2744, and in total they contain five BCGs. From these maps, we can also identify six other galaxies with sizeable GC populations in the outskirts of the maps. The distribution of bright GCs is not restricted to the main galaxy in the cluster, as they are significantly present throughout the intracluster medium. The lowest contour drawn in Fig.~\ref{fig:gcs-contours-all-blue-red} corresponds to the median number density of the image, indicating that \emph{most} of the identified bright star clusters lie between the galaxies in the cluster. Selecting only for complete GCs changes some of the details around the BCGs, as those areas have a higher local sky noise background, but it does not significantly change the overall distribution  of GCs around the cluster.

When comparing the blue and red subsamples (Fig.~\ref{fig:gcs-overlapping-blue-red}), we find that the blue GCs are more extended around galaxies and into the intergalactic medium than the red subsample, as observed in the local Universe \citep[e.g.][]{harris+2020}. The distribution of blue GCs shows a bridge connecting the north-western and northern clumps, indicating a past interaction between the two \citep[as suggested by][]{cha24}. Within that bridge, the red subsample shows an overdensity of objects, indicating the presence of one or more galaxies. There is also an overdensity of GCs between the main and the north-western clumps. The overdensity is present in both subsamples, but is more visible in the distribution of the red GCs because less of them lie in the intracluster medium.

\subsection{Predicted mass distributions from lensing models}


\begin{figure*}
    \centering
    \includegraphics[width=\hsize]{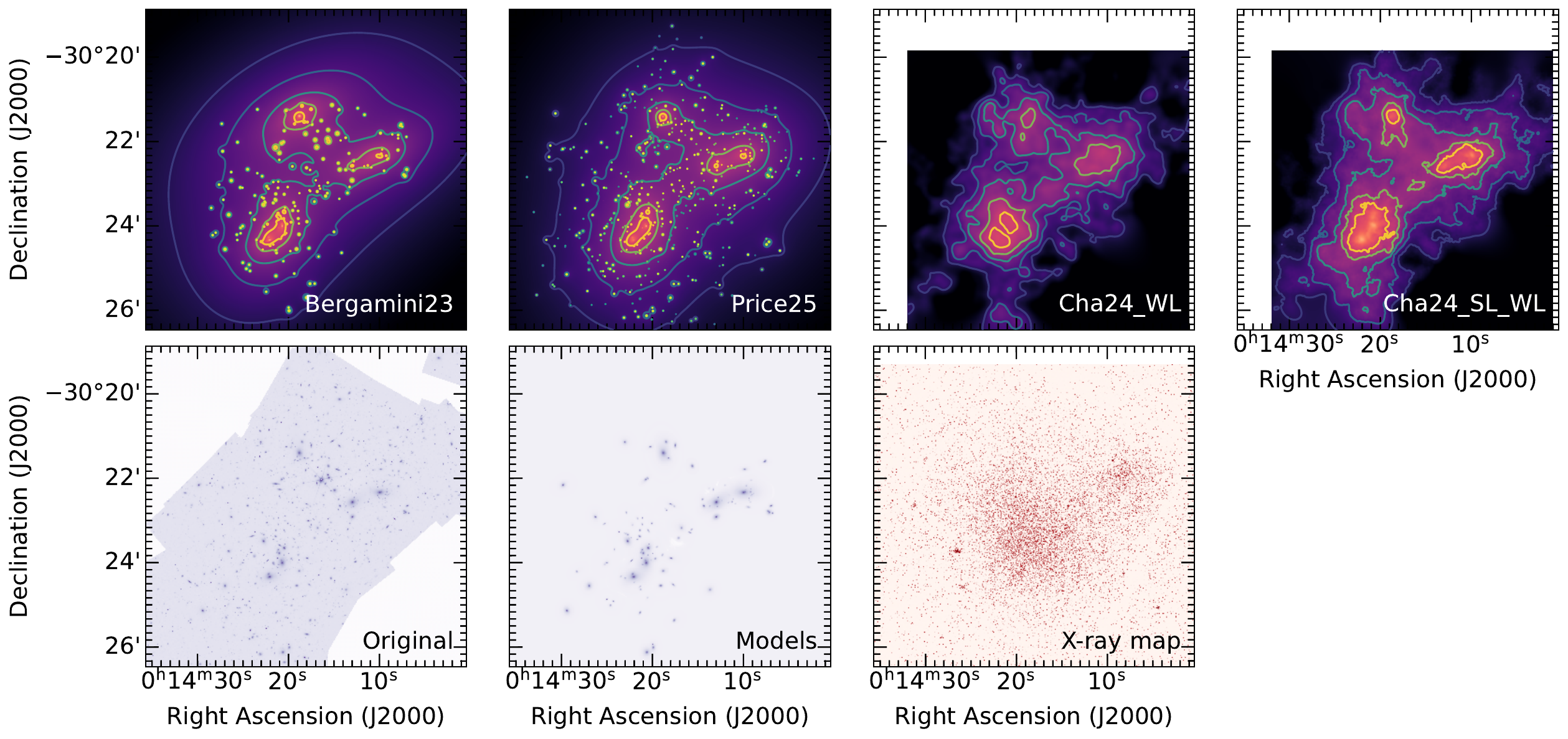}
    \caption{Multi-component view of Abell 2744. (Top row) Predicted mass distributions from strong and weak lensing analysis \citep[][respectively]{bergamini+2023, price25, cha24}. Contours are drawn at $\log_{10}(\Sigma/[\msun/\kpc^2]) = 8.3, 8.5, 8.7$, and $9.1$, and the colourbar is logarithmically normalized between $10^8$--$10^{10}~\msun/\kpc^2$. (Bottom left and middle panels) Original and modelled stellar light mosaics, averaged over the JWST/NIRCam LWC F277W, F356W and F444W from the DR3 of UNCOVER \citep{bezanson24,suess24,weaver24}. The colourbar shows the flux in logarithmic scale. (Bottom-right panel) Chandra $0.5$--$7.0$~keV broad band image showing the number of counts per pixel between $0$--$1$. All panels are shown within the same field of view as the `Price25' map. Regardless of the mass map, the galaxy cluster shows three prominent clumps that are connected between them. These clumps contain several large galaxies, and the large-scale X-ray emission is located in the center of the cluster, hinting of past interactions among the clumps.}
    \label{fig:lambda-maps}
\end{figure*}

Our goal is to identify the galactic component that most closely resembles the GC subsamples. For that, the first component we consider is the underlying mass distribution in the galaxy cluster, which we characterize with four different convergence maps obtained using a combination of strong and/or weak lensing constraints and spectroscopically-selected galaxies (see top row in Fig.~\ref{fig:lambda-maps}). \footnote{These maps correspond to a dimensionless surface density, $\kappa = \Sigma / \Sigma_{\rm cr}$, where the predicted mass surface density has been normalized by a critical surface density. We transform each map to the predicted mass surface density by multiplying them using $\Sigma_{\rm cr} = 1.020\times 10^9~{\rm M}_\odot/\rm kpc^{2}$ for the \citet{bergamini+2023} and \citet{price25} maps, and $1.777\times10^9~{\rm M}_\odot/\rm kpc^{2}$ for the \citet{cha24} maps.} We briefly describe here the assumptions and data used in each convergence map.

The first two convergence maps used assume that light traces the mass to model the masses of the cluster galaxies. The first one is the convergence map from \citet{bergamini+2023} (private communication). This map corresponds to a single realization of the best-fit lens model from the Monte-Carlo Markov chain. The strong lensing model uses a combination of imaging from the \textit{Hubble Space Telescope} (\hst), \textit{Magellan} and \jwst/NIRCam imaging, as well as spectroscopy from VIMOS, AAOmega, \hst, and VLT/MUSE. The parametric model is fit using the publicly available software LensTool with constraints from 149 multiple images from 50 background sources located $z\in[1.03, 9.76]$ and 669 spectroscopically-confirmed galaxies confirmed to be cluster members. The model uses non-truncated dual pseudo-isothermal elliptical
mass distributions (dPIEs) initially centered on the BCGs to describe the large-scale dark matter components. The novelty of this model is that, in addition to the two large-scale dark matter components initially centered on the two BCGs in the main cluster core, it also includes terms describing the cluster infalling region. It does so by adding two terms describing the cluster-scale dark matter components, one initially around the two BCGs in the north-west, and another initially around the BCG in the north. 

The second convergence map explored is part of the fourth data release of the UNCOVER collaboration \citep{furtak23, bezanson24, price25}. We use the best-fit model maps in $0.1''/\rm pix$ resolution of the v2.0 model available at \href{https://jwst-uncover.github.io/DR4.html}{https://jwst-uncover.github.io/DR4.html}. The strong lensing model uses an updated version of the analytic lens modelling code by \citet{zitrin15}. The model contains the compilation of spectroscopic cluster members and strong lensing features from \cite{bergamini+2023} in the main cluster core, and adds new multiple images from the UNCOVER imaging and spectroscopy in the northern and north-western DM structures. There are also more cluster member galaxies from BUFFALO \hst imaging \citep{steinhardt20} and Mega Science \jwst imaging \citep{suess24}. The parametric lens model includes 5 dPIEs cluster-scale DM halos centered on the five BCGs in the main cluster core and the northern and north-western sub-structures, and 552 cluster member galaxies. The model is constrained by 187 multiple images belonging to 66 background sources, 60 of those sources with spectroscopic redshifts.


The other two convergence maps correspond to the first mass reconstructions that do not assume that light-traces-mass and that they are profile-independent. These reconstructions include information on distortion signals (i.e.~weak lensing) and on multiple images (i.e.~strong lensing) \citep[][private communication]{cha24}. The reconstructions were made with the public \jwst/NIRCam imaging from the first release of the UNCOVER data \cite{bezanson24}. Although the data spans seven filters, the weak lensing distortions were detected only in F200W. To constrain the mass model, a total of 286 multiple images were used (136 of them with spectroscopically-confirmed redshifts), and the weak lensing distortions are measured with a source density of $350~{\rm arcmin}^{-2}$, the highest of any weak lensing study. 

The four convergence maps used cover a similar extent of the galaxy cluster, covering the inner $\sim30~\%$ of the virial radius of the galaxy cluster. Despite the similar coverage, the resolution of the maps is very different: the first two maps have $0.08''/\rm pix$ and $0.1''/\rm pix$, respectively, whereas the maps by \citet{cha24} have a coarser resolution of $1''/\rm pix$ (which correspond to $0.36, 0.46$, and $4.58~\rm kpc/pix$, respectively). To convert the convergence maps into predicted mass surface density maps, we multiply them by the following critical surface densities: $\Sigma_{\rm cr} = 1.020\times 10^9~\msun/\kpc^2$ for the maps of \citet{furtak23, bergamini+2023}, and $\Sigma_{\rm cr} = 1.777\times 10^9~\msun/\kpc^2$ for the maps of \citet{cha24}.  

\begin{figure*}
    \centering
    \includegraphics[width=\hsize]{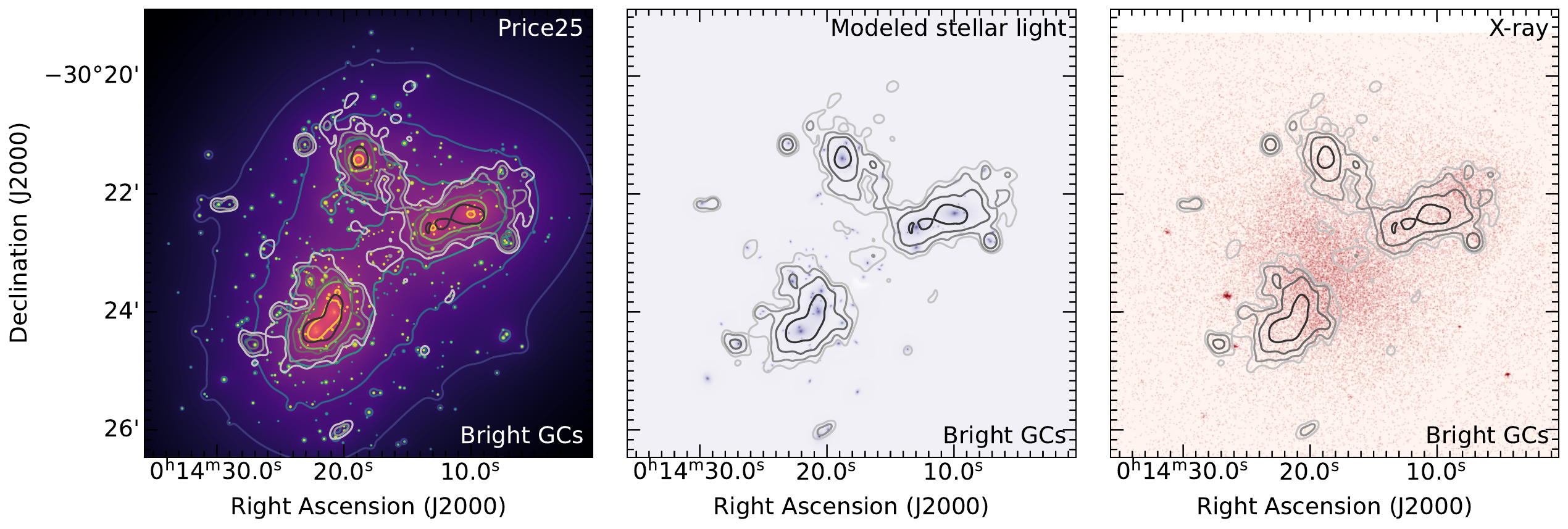}
    \caption{Spatial distribution of bright GCs (grey contours) overlaid over the projected total mass distributions (left-hand panel), the modelled stellar light mosaic from UNCOVER (middle panel) and the X-ray emission (right-hand panel) in Abell 2744. The grey contours in every panel represent the number density of GCs, weighted by their probability of recovery and smoothed by a Gaussian kernel. All panels are shown within the same field of view as the `Price25' map. Visually, the GCs contours closely trace the mass and stellar light maps, whereas they are offset relative to the X-ray emission.}
    \label{fig:mass-price25-gcs-contours}
\end{figure*}

\subsection{The stellar light}


The next component that we consider is the stellar population within the cluster, both in galaxies and in the intracluster medium, as viewed by the \jwst. For that, we use two sets of mosaics from the third data release by the UNCOVER team \citep{weaver24,bezanson24,suess24}\footnote{Available at \href{https://jwst-uncover.github.io/DR3.html}{https://jwst-uncover.github.io/DR3.html}}. The original mosaics provide a large-scale view of the galaxy cluster, which includes the galaxy members as well as foreground stars and background galaxies. In order to restrict the analysis to the galaxy members, we also use the stacked models of the bCGs. These models are iteratively produced by measuring the isophotal parameters of individual galaxies with \texttt{IRAF}, and they are the result of stacking over the 4 best models \citep[see][]{Shipley_2018, weaver24}.

The diffuse stellar light in the intracluster medium is expected to peak in redder bands due to it being composed of mainly an older stellar population \citep[e.g.][]{montes22}. Because of that, we create average maps combining the F277W, F356W and the F444W long wavelength channel (LWC) mosaics for both the original and the model maps. Although the modelled stellar light maps do not explicitly include the diffuse light emitted from the intracluster medium, we apply the same averaging to span a similar wavelength range. To produce the averaged maps, we first cut each LWC mosaic to the dimensions of the \citet{price25} convergence map and re-bin them by a factor of 2 to achieve a resolution of $0.08''$/pixel. We then add the resulting fluxes in the three channels to create the averaged map, and we offset the maps by the lowest value in the image to avoid negative values. That corresponds to $0.090796~[10~\rm nJy]$ and $0.004671~[10~\rm nJy]$ for the original map and the modeled stellar light, respectively. The final maps are shown in the bottom row in Fig.~\ref{fig:lambda-maps}.

\subsection{X-ray emission}

The last galactic component we examine is the hot, ionised gas, as traced by the emission of X-ray. For that, we retrieve the Advanced CCD Imaging Spectrometer (ACIS) image from the Chandra Data Archive \footnote{\href{https://cda.harvard.edu/chaser/}{https://cda.harvard.edu/chaser/}} for observation ID 8744 (PI Kempner)\footnote{\dataset [Chandra ObsId 8477]{https://doi.org/10.25574/08477}}. This observation was made in the $0.5$--$7.0$ keV broad band image with an exposure of $45.91$~ks, observed with ACIS-S in `VFAINT’ mode. The data was processed with CXC software using CalDB (version $8.4.4$). The resolution of the image is $1024$~pixels per side, with a spatial resolution of $0.5''$/pixel, and it represents counts per pixel. The image is shown in the bottom row of Fig.~\ref{fig:lambda-maps}. Whereas the mass and the stellar light images exhibit the five BCGs surrounded by smaller structures, the X-ray map shows emission largely from a large and smooth region located between the clumps. This would indicate that the X-ray emission traces the relic of a past interaction among the clumps in this galaxy cluster \citep[e.g.][]{owers+2011, medezinski16, rajpurohit21}.

\section{Results}\label{sec:results}

Here we present the comparison of the GC subsamples to the different galactic components. 

\subsection{Visual comparison of GCs to mass, stars and X-ray}

As a first step, we visually compare the spatial distribution of GCs in Abell 2744 to three representative maps of the mass distribution, the stellar light and the X-ray emission: the `Price25', the modeled stellar light from bCGs, and the X-ray emission maps. On top of these maps, we overlay contours describing the smoothed number density distributions of the sample of bright GCs. To create this map, we weight each GC by their probability of recovery, and apply a bi-dimensional Gaussian kernel of size $20~$kpc. The resulting figure is shown in Fig.~\ref{fig:mass-price25-gcs-contours}. 

Overall, we find very good agreement between the GC contours and the mass and the stellar light maps. As indicated above, the GCs are predominantly located around the three clumps of the galaxy cluster containing the five BCGs, and the smaller overdensities of GCs present in the outskirts of the cluster correspond to the GC populations in individual galaxies. These prominent GC systems in the outskirts only have corresponding subhaloes in the `Price25' mass map, highlighting the need of comprehensive and extensive datasets to develop the lensing model. 

The distribution of GCs around the main clump (towards the south-east of the cluster) reproduces the location of smaller galaxies scattered around the two BCGs. These are present in the `Bergamini23' and `Price25' mass maps because these models included profiles for spectroscopically-confirmed galaxies. The bridge of GCs present between the northern and north-western clumps is replicated in the mass distribution, and it contains galaxies visible in the stellar light map. 

In contrast, the distributions of GCs and of X-ray are mismatched; the X-ray emission mostly lies in the middle of the three clumps shown by the GC subsamples.

\subsection{Pixel-by-pixel correlations among maps}

\begin{figure}
    \centering
    \includegraphics[width=\hsize]{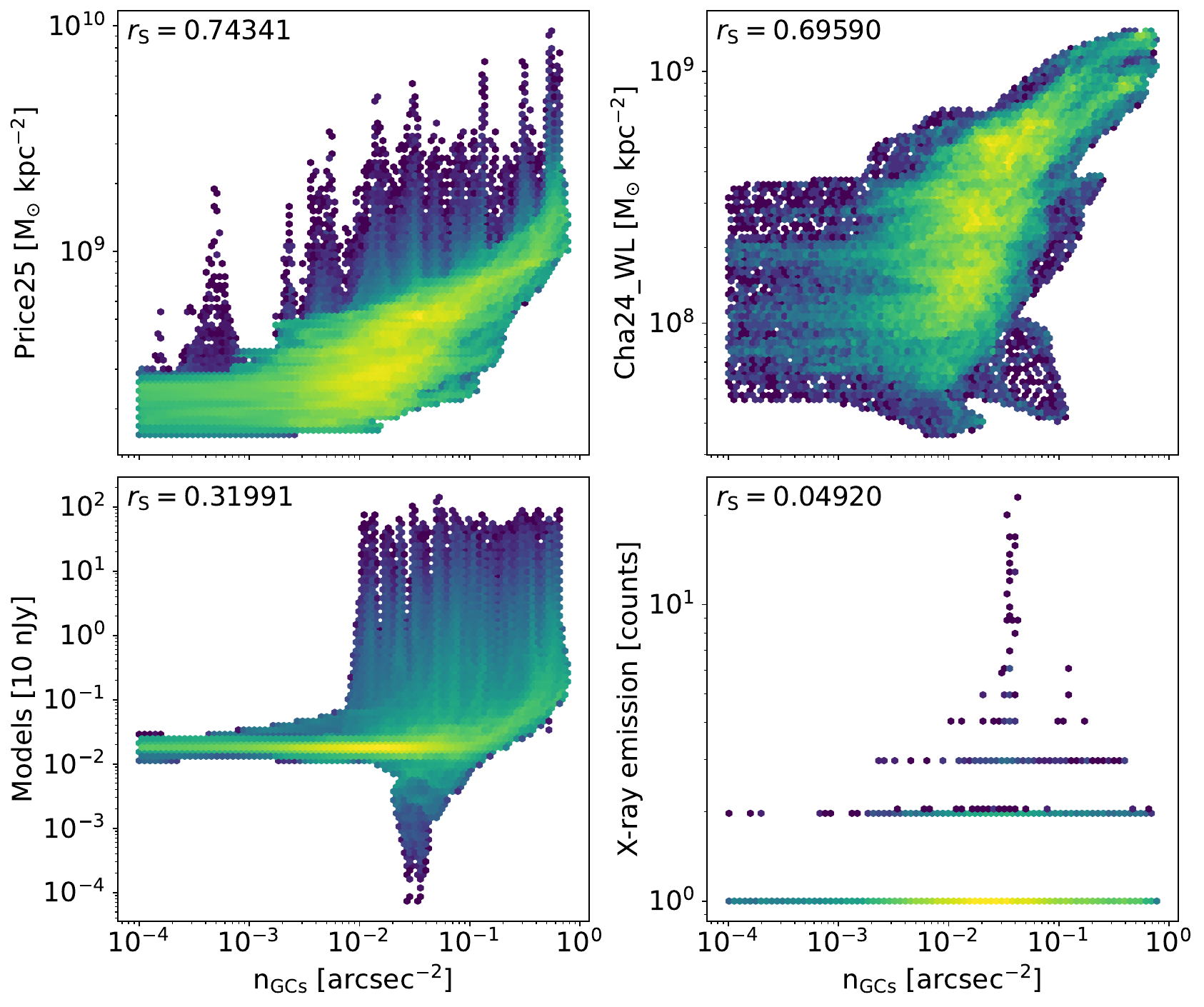}
    \caption{Pixel-by-pixel comparison between the smoothed spatial distribution of Bright GCs and different galactic components: the predicted mass maps from \cite{price25} and \cite{cha24} (top row), the modeled stellar light, and the X-ray emission. The Spearman rank coefficients are listed in the top-left corner of every panel. Bright GCs correlate more strongly with the mass maps than any other galactic component.}
    \label{fig:pixel-by-pixel-bright-gcs}
\end{figure}

We further compare the spatial distribution of GCs and that of different galactic components by looking at their pixel-by-pixel correlation (see Fig.~\ref{fig:pixel-by-pixel-bright-gcs}). To do this, we create number density maps of GCs that are weighted by the probabilities of recovery and smoothed by a bi-dimensional Gaussian kernel of size $20~$kpc. Importantly, these maps are at the same resolution and cover the same field of view as the map of the galactic component to be compared against. We then calculate the Spearman rank coefficient between these maps and the maps describing different galactic components for each GC subsample. 

We find that GCs correlate strongly with the mass distributions, regardless of the GC subsample or the mass map, with Spearman rank coefficients of $r_{\rm S} > 0.7$ for the Bright, Complete and Blue GC subsamples, and $r_{\rm S} > 0.6$ for the subsample of red clusters. As shown in Fig.~\ref{fig:gcs-contours-all-blue-red}, the redder star clusters are more strongly clumped around the main galaxies in the cluster, missing information on the intracluster medium. Defying expectations, the correlation between GCs and the modeled bCG light is a mere $r_{\rm S} = 0.3$. Two factors introduce scatter in the comparison and explain this low correlation coefficient. The first is the lack of an explicit description of the intracluster light in the models, meaning that pixels with high GC densities in between galaxies are paired with low flux values in the model image, thus flattening the trend. In addition, as in the mass maps, modeled galaxies without a corresponding observed GC overdensity introduce spikes of high stellar flux in regions of low GC densities (bottom-left panel in Fig.~\ref{fig:pixel-by-pixel-bright-gcs})\footnote{When repeating the analysis by smoothing both maps with the same Gaussian kernel, the scatter is greatly reduced because the small-scale features (i.e. the non-BCG galaxies in the cluster) disappear. Since the novelty of the `Bergamini23' and `Price25' maps is their high level of detail of the mass in the cluster, we show the correlations without smoothing them.}. In contrast, \citet{martis24} demonstrates that the correlation is stronger when the comparison is done against combined modelling of the BCGs and the intracluster light. The comparison of the GCs to the X-ray maps yield the lowest Spearman rank coefficients, with $r_{\rm S} < 0.1$ (bottom-right panel in Fig.~\ref{fig:pixel-by-pixel-bright-gcs}), indicating that there is no correlation between the location of GCs and that of X-ray emitting gas, as expected from Fig.~\ref{fig:mass-price25-gcs-contours}. By doing this pixel-by-pixel comparison, we find that GCs are a better tracers of mass than of any other galactic component, but the Spearman rank correlation coefficient cannot discriminate between the mass maps. 

\begin{figure*}
    \centering
    \includegraphics[width=\hsize]{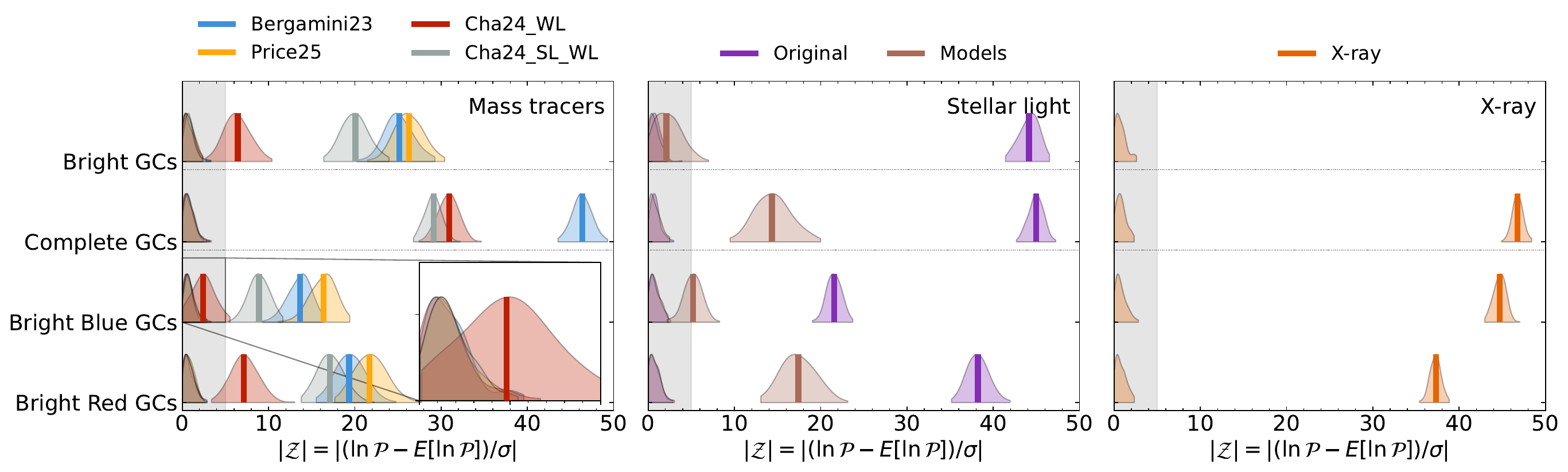}
    \caption{Comparison of the $\mathcal{Z}$--scores between our samples of GCs and maps of different galactic components. This score is anchored by the mean and standard deviation, $E[\ln \mathcal{P}]$ and $\sigma$, of the expected distribution of probability densities of a given map ($\ln\mathcal{P} \{ (B_i)^{\lambda_{\rm eff}} | \lambda_{\rm eff} \}$, violin plots around $0$). Vertical lines show the $\mathcal{Z}$--score of the GC sample against a galactic component, and the violin plots correspond to the bootstrapped distributions. The grey shaded region shows the area of compatibility. The Bright Blue GCs are in good agreement with the `Cha24\_WL' map (overlap is better seen in the inset).}
    \label{fig:zscore-allgc-samples}
\end{figure*}

\subsection{Log-likelihoods assuming an inhomogenous Poisson point process}\label{sec:probabilities}

In order to discriminate to which map (if any), GCs are better tracers of, we apply the statistical method based on assuming an inhomogenous spatial Poisson process (see Sect.~\ref{sec:methods}). For this, we calculate the log-likelihood of the spatial distribution of GCs having been spawned from that of another galactic component. To do this calculation, we use the actual locations of the GCs within the galaxy cluster, without applying any smoothing. We compare these empirical log-likelihoods to the expected distribution of log-likelihoods for a given map via the $\mathcal{Z}$--score, which normalizes the log-likelihood by their expected value and standard deviation. Thus, this metric effectively indicates how many standard deviations the results from the GC samples are relative to the self-map comparison. We do so for our four GC samples, which we show in Fig.~\ref{fig:zscore-allgc-samples}. The quoted uncertainties correspond to the $95\%$ confidence interval from the bootstrapped distributions.

Focusing first on the sample of Bright GCs, we find that the 
$\mathcal{Z}$--scores against the mass tracers are the lowest ($6 \pm 3 < |\mathcal{Z}| < 26 \pm 3$), whereas the $\mathcal{Z}$--score against the X-ray map is out of scale ($57\pm 1$, respectively). Among the mass maps, the lowest score is against the `Cha24\_WL' map, which is derived only from weak lensing constraints. Interestingly, this agreement is also found for the other GC samples, suggesting that GCs provide similar constraints pn the underlying mass distribution as do weak lensing signals. There is also close agreement with the modeled stellar light, with $|\mathcal{Z}| = 2 \pm 3$. In contrast, the $\mathcal{Z}$--score against the stellar light map is high, $\mathcal{Z} = 44\pm 2$. Despite having averaged over three red filters, that image contains everything within the field of view; i.e.~foreground stars, galaxies in the cluster and in the background. In order to properly assess the correlation between GCs and diffuse light in the cluster, this result highlights the need for detailed modeling of the intracluster light within the galaxy cluster \citep[e.g.][]{montes_trujillo2014}.

In contrast, the sample of complete GCs has the highest $\mathcal{Z}$--scores against the mass tracers among our four samples. Although the smoothed spatial distribution differs only slightly from that of the Bright GCs (see Fig.~\ref{fig:gcs-contours-all-blue-red}), in reality the low local sky noise criterion defining this sample excludes the centers of galaxies. This implies that, in order to reconstruct the mass distribution, tracers are needed both in the intracluster medium and around galactic centers.

When separated based on color, the bluer sample shows the lowest $\mathcal{Z}$--scores against the mass maps amongst all the samples. Compared to the red subsample, the blue GCs have lower $\mathcal{Z}$--scores by a factor of $1.5$--$3$. This is an expected result, since bluer GCs are scattered further into the intracluster medium. In particular, the sample of blue GCs lies within the expected distribution of log-likelihoods for the `Cha24\_WL' map (with $|\mathcal{Z}| = 2.4\pm2.5$, see the inset), indicating that the spatial distribution of bluer GCs is in agreement with the mass as traced by weak lensing. 

Regardless of the GC sample, the scores against the `Bergamini23' and `Price25' mass maps are always the highest. Those maps are derived from a combination of multiple images from strong gravitational lensing and spectroscopically-confirmed galaxies, which render very detailed high-resolution maps (see Fig.~\ref{fig:lambda-maps}). Despite our large sample of GCs, most of the smaller haloes introduced from the spectroscopically-selected galaxies lack an observed GC system (Fig.~\ref{fig:mass-price25-gcs-contours}), which affects the comparison.

\subsection{Cross-map comparisons}

After assessing that GCs are more closely correlated to the mass, we can now invert our question and examine which of the galactic components better traces each other. For that, we perform cross-map comparisons as described in Sect.~\ref{subsec:cross-map}, where we spawn data points from a map $\lambda_1$ (i.e., `tracer' map) and compute the log-likelihood against a map $\lambda_2$ (i.e., the map being traced). 


It is worth noting that the cross-comparisons of maps do not yield symmetric results. The asymmetry is informative as it encodes morphological information about the two maps. An insightful example is to examine the cross-comparisons between extended and compact distributions. The $\mathcal{Z}$--score obtained when spawning from an extended distribution and comparing against a compact one is much larger relative to the compact-compact comparison than when doing the opposite experiment. That is because most spawned points from the extended distribution will lie in regions of the compact map with little signal. In contrast, the spawned points from a compact distribution will always lie in the central region of the extended distribution. Since the log-likelihood is a summation over all the spawned points, it thus heavily relies on the contribution of each one. Because of this, we require symmetry to interpret the $|\mathcal{Z}|$--scores.

\begin{figure*}
    \centering
    \includegraphics[width=\hsize]{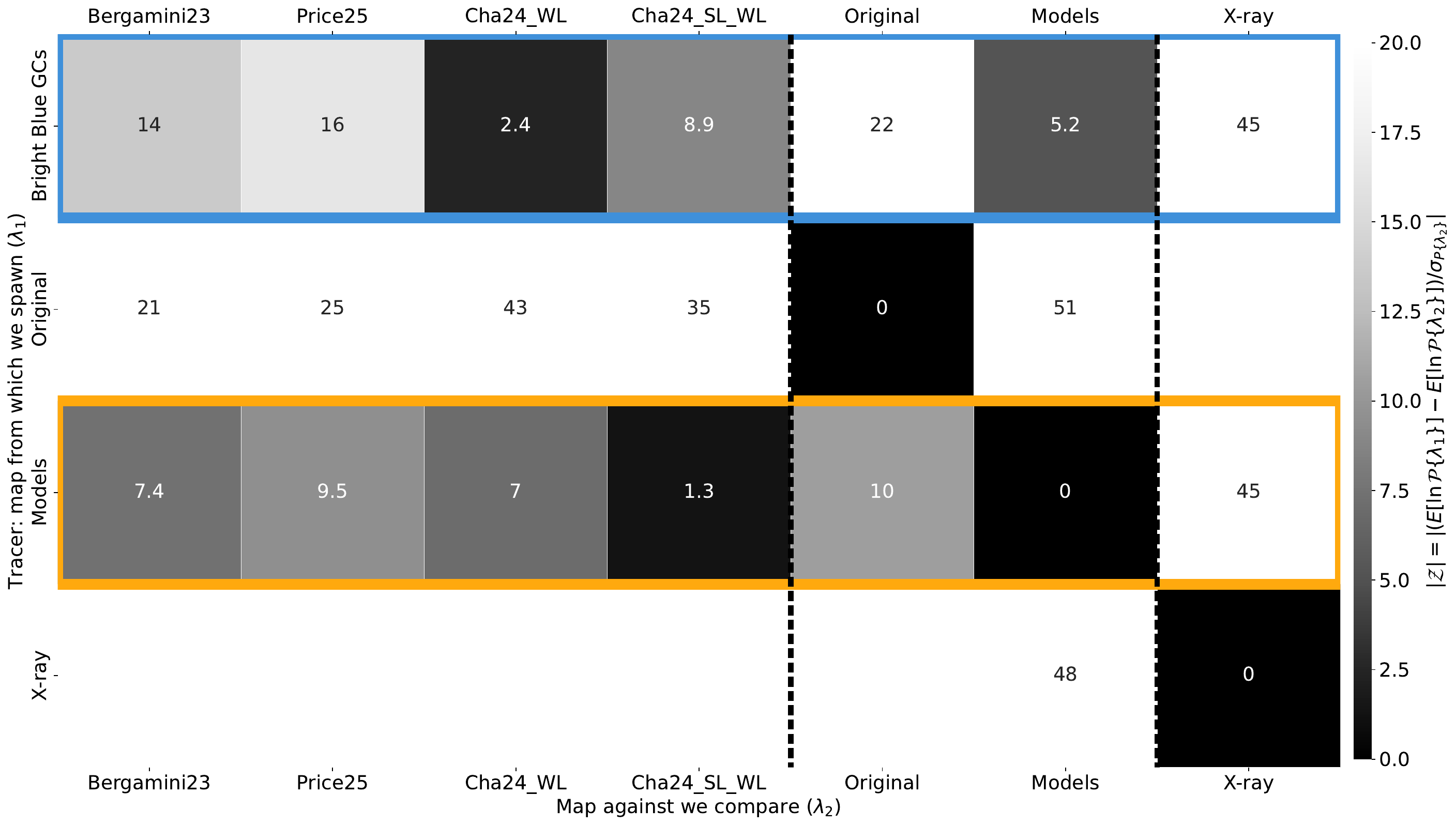}
    \caption{Cross comparison among some of the galactic components of Abell 2744 considered in this work. The colorbar corresponds to the median $\mathcal{Z}$-score of the log-likelihoods: spawning datapoints from the tracer map $\lambda_1$ (vertical axis) and calculating the log-likelihood against $\lambda_2$ (horizontal axis), $\ln\mathcal{P} \{ (x_i , y_i, \pmb{\theta}_i)^{\lambda_1} | \lambda_{2} \}$. Darker colors indicate a closer agreement between the maps, and masked cells correspond to asymmetric comparisons. The blue and orange horizontal boxes highlight the role of Bright Blue GCs and the modeled stellar light as tracers, and vertical black dotted lines divide the horizontal axis in the type of galactic component these maps represent.}
    \label{fig:zscore-cross-all-models}
\end{figure*}

This morphological case help us interpret the cross-maps comparisons shown in Fig.~\ref{fig:zscore-cross-all-models} (masked cells are asymmetric comparisons). Rows in this figure show the median $\mathcal{Z}$--score of a fixed `tracer' map ($\lambda_1$) given different maps $\lambda_2$, whereas columns hold the values for a fixed map being traced given different tracers. For simplicity, we restrict the discussion to the case when the tracer maps are observables (i.e.~GCs, stellar light or X-ray emission).

Somewhat unexpectedly, we find that the best tracer of the `Price25' map (and other convergence maps) is the X-ray emission map. However, the opposite is not true. Following the morphological case discussed above, this is due to the X-ray emission being more compact and centered between the three sub-clumps, whereas any other distribution is more extended. Thus, this is an spurious agreement and it is masked. 

Amongst the tracer maps, the modeled stellar light map (orange rectangle in Fig.~\ref{fig:zscore-cross-all-models}) consistently shows lower $\mathcal{Z}$--scores against the mass maps compared to any other galactic component, except for the Blue GCs against the `Cha24\_WL' map (blue rectangle). This implies that detailed modelling of the galaxies in the cluster holds most of the information about the main centers of mass in the cluster. We note that the intracluster medium is missing from this modelling, and the comparison might improve in future models that account for this component. In contrast, the original mosaics of the galaxy cluster are heavily contaminated with non-cluster sources, and they show the worst agreement against any tracer map.


\section{Discussion and conclusions}\label{sec:conclusions}

In this article, we present a statistical method to determine which galactic component better reproduces the spatial distribution of GCs in a galaxy cluster. For that, we assume that the spatial location of GCs follows an inhomogeneous Poisson point process modulated by the galactic component considered, and we apply the methodology to the GCs in Abell 2744. This method allows us to discriminate to which galactic component GCs are better tracers of, and to examine the cross-relations between the different components.

Dividing the GC catalogue in Abell 2744 in four samples (Fig.~\ref{fig:gcs-contours-all-blue-red}), we find that bright GCs reside mainly around the BCGs and large galaxies in the cluster, with a substantial population also scattered in the intracluster medium. Restricting the GC sample to those in low-sky regions (Zones 1 and 2) leads to a lack of GCs near the centers of galaxies, as those environments have higher local sky noise, which decreases the probability of recovery of the GCs. at faint magnitudes. When dividing based on color, we find that the intracluster medium is mainly populated by blue GCs (Fig.~\ref{fig:gcs-overlapping-blue-red}). Interestingly, a clear extended overdensity in the distribution of red GCs is visible between the north and north-eastern sub-clumps, with a compact overdensity of GCs present in all the samples. This suggests a previous interaction between these sub-clumps. 

We find that the spatial distribution of Bright GCs qualitatively agrees with that of the predicted mass maps and BCG light, but not with the X-ray emission (Fig.~\ref{fig:mass-price25-gcs-contours}). We quantify this result by determining the Spearman rank coefficient in a pixel-by-pixel analysis of the images, and we find that the coefficient is higher than $r_{\rm S} > 0.7$ only when comparing against the mass maps (Fig.~\ref{fig:pixel-by-pixel-bright-gcs}). However, this method cannot distinguish to which map GCs are better correlated against.

For this, we apply the methodology derived in Sect.~\ref{sec:methods} to compare our samples of GCs against maps of different galactic components. For each comparison, we determine its $|\mathcal{Z}|$, which is a measure of how far away the log-likelihood for the GC sample is relative to the expected distribution of log-likelihoods for that galactic component. Regardless of the GC sample, we find that the comparisons with the mass maps yield the lowest $|\mathcal{Z}|$--scores. 

In particular, the Bright Blue GCs sample has a $|\mathcal{Z}|$--score compatible with the expected distribution from the `Cha24\_WL' (Figs.~\ref{fig:zscore-allgc-samples} and \ref{fig:zscore-cross-all-models}). This suggests that the constraints on the mass distributions provided by GCs are of similar precision to those derived from weak lensing. The agreement is worse for mass maps including constraints from strong gravitational lensing and spectroscopically-selected galaxies, because there are no detected bright clusters in the majority of those galaxies. Future deeper catalogues will provide information of GC systems in a wider region of the galaxy cluster.


The wide, deep mosaics of lensing galaxy clusters taken with \jwst/NIRCam are of exceptional quality enabling deep photometry of their GC populations, and allowing us to use them to trace the underlying mass distribution. Recent studies in relaxed galaxy clusters focus on the number density radial profile of GCs, and they correlate it against the stellar light and the lensing map of the cluster \citep[e.g.][]{diego23,diego24b,martis24,diego26b,diego26a}. Although this technique is easy to model and interpret in relaxed systems, non-virialized galaxy clusters with multiple BCGs (like Abell 2744) present a much more complex scenario. To overcome this, the statistical method developed in this paper offers the flexibility to compare the GC distribution against continuous maps describing galactic components in any type of environment. 

The method presented here is readily available to be applied on other galaxy clusters with similar datasets. The code is publicly available on GitHub\footnote{\href{https://github.com/mreinacampos/starclusters-in-jwst}{https://github.com/mreinacampos/starclusters-in-jwst}}. In future papers, we will examine the correlation between GCs and the mass distribution within lensing clusters over a wider range of redshifts, between $0.2<z<1$.

\begin{acknowledgments}
MRC thanks Sangjun Cha and Pietro Bergamini for sharing the convergence maps of their lensing models, and John Weaver and Rachel Bezanson for help navigating the UNCOVER data.

As with \citetalias{harris_reinacampos2023,harris_reinacampos2024}, we acknowledge the work of the UNCOVER team to produce the beautiful mosaic images used in this work. These images for Abell 2744 are publicly available at: \href{https://jwst-uncover.github.io/#}{https://jwst-uncover.github.io/\#}.

MRC gratefully acknowledges the Canadian Institute for Theoretical Astrophysics (CITA) National Fellowship for partial support, and the Global Talent Junior Fellowship Programme at the Instituto Galego de F\'isica de Altas Enerx\'ias (IGFAE), supported by grant CEX2023-001318-M (Mar\'ia de Maeztu Unit of Excellence and Agencia Española de Investigaci\'on / 10.13039/501100011033). This work was supported by the Natural Sciences and Engineering Research Council of Canada (NSERC) [funding reference number 568580], and by the Xunta de Galicia (CIGUS Network of Research Centres) and the European Union through the Galicia Feder 2021-2027 Program. JSS was supported by funding from NSERC Discovery Grant RGPIN-2023-04849.

The scientific results reported in this article are based in part on data obtained from the Chandra Data Archive (ObsId8477).

\end{acknowledgments}

\begin{contribution}



All authors came up with the initial research concept and edited and reviewed the manuscript. MRC lead the project: obtained the data, wrote the software and the manuscript, and administers the project GitHub repository. JSS provided the formal statistical analysis, and WEH supervised the analysis.

\end{contribution}

%

\software{Astropy \citep{astropy:2013, astropy:2018, astropy:2022},
        marimo \citep{Agrawal_marimo_-_an_2023},
          Matplotlib \citep{hunter2007},
          Numpy \citep{harris+2020b},
          pandas \citep{pandas_allversions},
          Scipy \citep{Jones01},
          Seaborn \citep{Waskom2021}}
    
\dataset[Chandra ObsId 8477]{https://doi.org/10.25574/08477}


\appendix

\section{Robustness of the statistical method}\label{app:robustness}

To test the robustness of the statistical method presented in this paper, we summarise in Tab.~\ref{tab:se-means} the means (or expected value) and standard deviations of the log-likelihoods distributions for the self-map comparisons presented as violin plots in Fig.~\ref{fig:zscore-allgc-samples}.

\begin{table*}
\centering
\caption{Mean and standard deviation of the distributions of $\ln\mathcal{P}$ for the self-map comparison of every galactic component, with their respective standard errors.}
\label{tab:lnP_stats}
\setlength{\tabcolsep}{4pt}
\begin{tabular}{l cc cc cc cc}
\hline\hline
 & \multicolumn{2}{c}{Bright GCs}
 & \multicolumn{2}{c}{Complete GCs}
 & \multicolumn{2}{c}{Bright Blue GCs}
 & \multicolumn{2}{c}{Bright Red GCs} \\
\cmidrule(lr){2-3}\cmidrule(lr){4-5}\cmidrule(lr){6-7}\cmidrule(lr){8-9}
Model
 & $E[\ln(\mathcal{P})]$ & $\sigma$
 & $E[\ln(\mathcal{P})]$ & $\sigma$
 & $E[\ln(\mathcal{P})]$ & $\sigma$
 & $E[\ln(\mathcal{P})]$ & $\sigma$ \\
\hline
\textsc{Price25}
 & $-22156 \pm 4$ & $42 \pm 3$
 & $-18077 \pm 4$ & $36 \pm 3$
 & $-12078 \pm 3$ & $30 \pm 2$
 & $-10080 \pm 3$ & $28 \pm 2$ \\
\textsc{Cha24\_WL}
 & $-9739 \pm 6$  & $55 \pm 4$
 & $-7956 \pm 5$  & $49 \pm 3$
 & $-5313 \pm 4$  & $40 \pm 3$
 & $-4442 \pm 4$  & $39 \pm 3$ \\
\textsc{Cha24\_SL\_WL}
 & $-15315 \pm 6$ & $56 \pm 4$
 & $-12488 \pm 7$ & $66 \pm 5$
 & $-8348 \pm 5$  & $48 \pm 3$
 & $-6973 \pm 4$  & $41 \pm 3$ \\
\textsc{Original}
 & $-49225 \pm 5$ & $49 \pm 3$
 & $-40159 \pm 4$ & $44 \pm 3$
 & $-26837 \pm 4$ & $41 \pm 3$
 & $-22401 \pm 3$ & $34 \pm 2$ \\
\textsc{Models}
 & $-57949 \pm 7$ & $67 \pm 5$
 & $-47254 \pm 5$ & $53 \pm 4$
 & $-32121 \pm 9$ & $88 \pm 6$
 & $-26361 \pm 5$ & $49 \pm 3$ \\
\textsc{X-ray}
 & $-109790 \pm 97$ & $970 \pm 69$
 & $-89774 \pm 93$ & $931 \pm 66$
 & $-59891 \pm 67$ & $671 \pm 48$
 & $-50118 \pm 68$ & $676 \pm 48$ \\
\hline
\multicolumn{9}{c}{$\mu \pm SE_\mu$ and $\sigma \pm SE_\sigma$ of the log-likelihood $\ln\mathcal{P}$ distributions, where $SE_\mu = \sigma/\sqrt{n}$ and $SE_\sigma = \sigma/\sqrt{2(n-1)}$. }
\end{tabular}\label{tab:se-means}
\end{table*}


\bibliography{mybib}{}
\bibliographystyle{aasjournalv7}



\end{document}